\documentclass[aps,prb,showpacs,reprint,superscriptaddress]{revtex4-1}
\usepackage[utf8]{inputenc}
\usepackage[pdftex]{graphicx}
\usepackage{amssymb}
\usepackage{physics}
\usepackage{amsmath}
\usepackage{xfrac}
\usepackage{graphicx}
\usepackage{dcolumn}
\usepackage{color}
\usepackage{bm}
\usepackage{hyperref}

\newcommand{\ymb}{YbMnBi$_2$}
\graphicspath{{./}}
\begin{document}
\preprint{APS/123-QED}
\title{Magnetic structure and excitations of the topological semimetal \ymb}
\author{Jian-Rui Soh}
\affiliation{Department of Physics, University of Oxford, Clarendon Laboratory, Parks Road, Oxford OX1 3PU, UK}
\author{Henrik Jacobsen}
\affiliation{Department of Physics, University of Oxford, Clarendon Laboratory, Parks Road, Oxford OX1 3PU, UK}
\author{Bachir Ouladdiaf}
\affiliation{Institut Laue-Langevin, 6 rue Jules Horowitz, 38042 Grenoble Cedex 9, France}
\author{Alexandre Ivanov}
\affiliation{Institut Laue-Langevin, 6 rue Jules Horowitz, 38042 Grenoble Cedex 9, France}
\author{Andrea Piovano}
\affiliation{Institut Laue-Langevin, 6 rue Jules Horowitz, 38042 Grenoble Cedex 9, France}
\author{Tim Tejsner}
\affiliation{Institut Laue-Langevin, 6 rue Jules Horowitz, 38042 Grenoble Cedex 9, France}
\affiliation{Nanoscience Center, Niels Bohr Institute, University of Copenhagen, DK-2100 Copenhagen, Denmark}
\author{Zili Feng}
\affiliation{Beijing National Laboratory for Condensed Matter Physics, Institute of Physics, Chinese Academy of Sciences, Beijing 100190, China}
	\author{Hongyuan Wang}
	\affiliation{School of Physical Science and Technology, ShanghaiTech University, Shanghai 201210, China}
	\affiliation{University of Chinese Academy of Sciences, Beijing 100049, China}
	\author{Hao Su}
	\affiliation{School of Physical Science and Technology, ShanghaiTech University, Shanghai 201210, China}
	\author{Yanfeng Guo}
	\affiliation{School of Physical Science and Technology, ShanghaiTech University, Shanghai 201210, China}
	\author{Youguo Shi}
	\affiliation{Beijing National Laboratory for Condensed Matter Physics, Institute of Physics, Chinese Academy of Sciences, Beijing 100190, China}
	\author{Andrew T. Boothroyd}
	\email{andrew.boothroyd@physics.ox.ac.uk}
	\affiliation{Department of Physics, University of Oxford, Clarendon Laboratory, Parks Road, Oxford OX1 3PU, UK}%
	\date{\today}
	%May be coupled but not canted, We are saying not canted and not coupled
	\begin{abstract}
		We investigated the magnetic structure and dynamics of \ymb, with elastic and inelastic neutron scattering, to shed light on the topological nature of the charge carriers in the antiferromagnetic phase.  We confirm C-type antiferromagnetic ordering of the Mn spins below $T_{\rm N} = 290$\,K, and determine that the spins point along the $c$-axis to within about $3^\circ$.
		%we rule out the predicted spontaneous $10^\circ$ canting of the Mn moments away from the crystal $c$-axis and put an upper limit on the tilt of about $3^\circ$. 
		The observed magnon spectrum can be described very well by the same effective spin Hamiltonian as was used previously to model the magnon spectrum of CaMnBi$_2$.
		%Furthermore, the magnetic excitation spectrum of the ordered Mn$^{2+}$ moments does not display any anomalous behavior, indicative of a strong coupling between the AFM order and the predicted Weyl fermions.} 
		Our results show conclusively that the creation of Weyl nodes in YbMnBi$_2$ by the time-reversal-symmetry breaking mechanism can be excluded in the bulk.
	\end{abstract}
	
	%75.25.-j spin arrangements
	%75.30.Ds Spin waves
	%75.30.Gw Magnetic anisotropy
	%74.70.Xa Pnictides and Chacogenides
	
	\maketitle
	\section{introduction \label{sec:level1}}
	% https://arxiv.org/pdf/1605.04199.pdf
	Dirac and Weyl materials are semimetals whose valence and conduction bands have a linear dispersion in the vicinity of the Fermi energy~\cite{Burkov2016,Armitage2018}. These gapless band crossings, which are protected by topology or crystalline symmetries, can give rise to massless quasi--particle excitations which can be described by the relativistic Dirac or Weyl equations. Materials that host such fermions possess a range of desirable physical properties: exceptionally high electrical and thermal conductivities, immunity to disorder and ballistic electronic transport~\cite{Rau2016,Pesin2009,Hasan2010}. 
	
	Weyl semimetals (WSMs) can occur in crystals with broken spatial inversion symmetry (IS), broken time-reversal symmetry (TRS), or both. Examples of the first type (with broken IS only) were found in 2015~\cite{Lv2015,Huang2015,Xu2015,Yang2015}, but realizations of WSMs with broken TRS are still rare~\cite{Armitage2018}. Recently, the layered AFM \ymb\, was proposed as a potential candidate~\cite{Borisenko2015}. The evidence from angle-resolved photoemission spectroscopy (ARPES) is quite convincing~\cite{Borisenko2015}, and there is also some support from optics~\cite{Chaudhuri2017,Chinotti2016}. 
	
\begin{figure}[t]
		\includegraphics[width=0.5\textwidth]{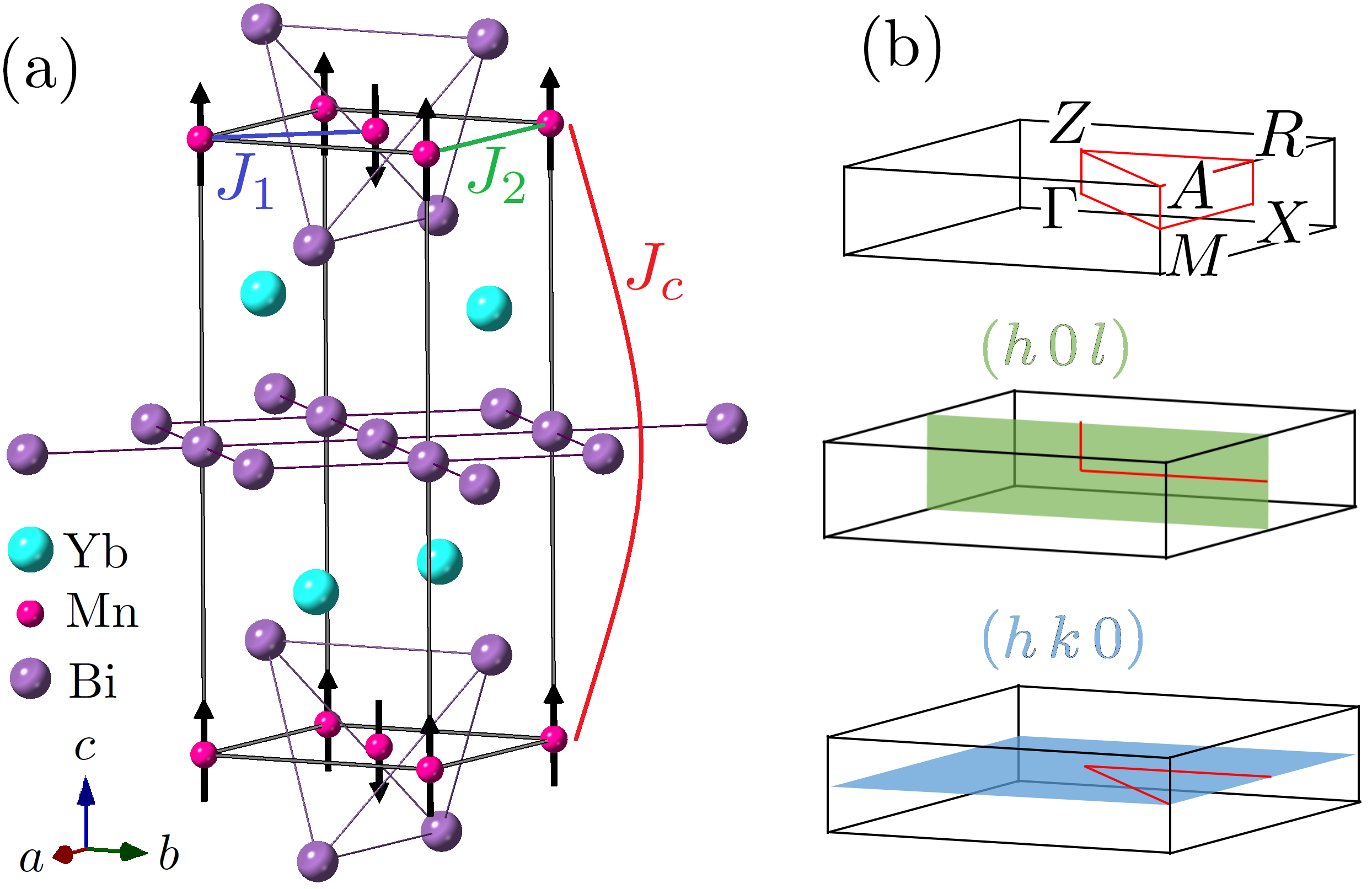}
		\caption{\label{fig:Unitcell} (a) The unit cell of \ymb\, for the space group $P4/nmm$ (No.~129). The proposed Weyl fermions are contained in the Bi square net in the center of the unit cell. The magnetic exchange between the $ab$-plane nearest neighbor ($J_1$), $ab$-plane next-nearest neighbor ($J_2$), and $c$-axis nearest neighbor ($J_c$) Mn$^{2+}$ ions were used in the linear spin-wave model to describe the magnon spectrum. (b) The definition of high symmetry lines and planes in the first Brillouin zone of the tetragonal lattice. The spin-wave spectrum in the $(h0l)$ and $(hk0)$ reciprocal lattice planes was mapped in this work. Here, the reciprocal lattice vector is defined as, $\mathbf{G} = h \mathbf{b}_1 + k \mathbf{b}_2 + l \mathbf{b}_3$, where $|\mathbf{b}_1| = |\mathbf{b}_2| = 2\pi/a$ and $|\mathbf{b}_3| = 2\pi/c$.}
	\end{figure}

	The tetragonal unit cell of \ymb, which can be described by the $P4/nmm$ space group  (No.~129), includes alternating Bi square layers that host the possible Weyl fermions,~\cite{Klemenz2019,Borisenko2015,Chinotti2016,Chaudhuri2017,Wang2016,Liu2017,Pal2018} and MnBi$_4$ tetrahedral layers which contain magnetic moments on the Mn atoms [See Fig.~\ref{fig:Unitcell}(a)]. In the antiferromagnetically (AFM) ordered phase, below $T_\textrm{N} = 290$\,K, neighbouring Mn spins are reported to be antiparallel within the $ab$ plane, but crucially, they are ferromagnetically stacked along the $c$--axis~\cite{Wang2016,Zaliznyak2017,Liu2017}. This means that magnetic coupling to the Bi conduction states is allowed at the mean--field level, which can lead to band splitting. 
	
	In Ref.~\onlinecite{Borisenko2015}, it was argued that creation of Weyl points by TRS breaking in \ymb\, requires a $\sim$10$^\circ$ canting of the Mn moments away from the  $c$-axis. If present, this canting would generate a net ferromagnetic component in the $ab$-plane of \ymb\,, and would account for the Weyl nodes and arcs observed in the ARPES data. Such a small deviation in the moment direction from the $c$-axis would not have been discernible in the $(100)$ magnetic peak studied in the previous neutron diffraction measurements~\cite{Wang2016,Zaliznyak2017,Liu2017}, so the possibility that \ymb\, might be a WSM by this mechanism remains to tested.
	
	Moreover, if the AFM order of manganese creates Weyl fermions, which then dominate the electronic transport~\cite{Wang2016,Liu2017}, then these quasiparticle excitations could play be expected to play some role in the exchange coupling between Mn moments which could in turn influence the magnon spectrum. 
	%For example, the coupling between moments along the $c$-axis might be anomalously large.
	As the magnetic order is key to the behavior of YbMnBi$_2$ as a topological material, measurements of the magnon spectrum, and the exchange parameters derived from it, could provide additional information on the presence of Weyl fermions near the Fermi energy.
	
	In light of this, we set out in this study, (i) to search for evidence of a canted magnetic structure by neutron diffraction, and (ii) to investigate the magnon spectrum in the AFM phase of \ymb\, through inelastic neutron scattering. To achieve the required sensitivity to the predicted ferromagnetic component of the proposed canted magnetic structure, we performed careful measurements of the weak $(00l)$ nuclear reflections. Furthermore, to identify any anomalies in the magnetic exchange between Mn moments associated with the presence of Weyl fermions, we compare the observed magnon spectrum with that of Dirac semimetal CaMnBi$_2$~\cite{Rahn2017}, which is isostructural to \ymb. We demonstrate that the Mn sublattice in \ymb\, has C-type AFM ordering below $T_{\rm N} = 290$\,K, with the moments aligned along the $c$-axis to within  $3^\circ$ (at $95\%$ confidence level). Moreover, we find no evidence from the magnon spectrum for anomalous magnetic coupling between the Mn spins. Our results rule out the existence of magnetically-induced Weyl fermions in the bulk of \ymb\,, but leave open the possibility that the $\sim$10$^\circ$ canting of the Mn moments needed to form the Weyl nodes might occur at the surface.
	
	\section{Experimental Details}
	Single crystalline \ymb\, was grown by the self-flux method. The starting materials were mixed together in a molar ratio of Yb:Mn:Bi = 1:1:8. The mixture was placed into an alumina crucible, sealed in a quartz tube, then slowly heated to 900$^\circ$C and kept at this temperature for 10 hours. The assembly was subsequently cooled down to 400$^\circ$C at a rate of 3$^\circ$C/hour. It was finally taken out of the furnace at 400$^\circ$C and was put into a centrifuge immediately to remove the excess Bi. The structure and quality of the single crystals was checked with laboratory x--rays on a 6--circle diffractometer (Oxford Diffraction) and Laue diffractometer (Photonic Science). A superconducting quantum interference device (SQUID) magnetometer (Quantum Design) was used to study the magnetization of \ymb\, as a function of temperature. These zero-field-cooled (ZFC) magnetometry measurements were performed in the temperature range 10 to 370\,K in a  field of 1\,T applied parallel to the $a$- and $c$-axes of YbMnBi$_2$. 
	
	Elastic neutron scattering of a \ymb\, single crystal with a mass of 76 mg was performed on a 4--circle diffractometer (D10) at the Institut Laue-Langevin (ILL) reactor source. The intensities of the reflections were studied over the temperature range of 20 to 400\,K. A pyrolytic graphite (PG) monochromator was used to select the incident neutron wavelength of $\lambda = 2.36$\,\AA . The rocking curve of each peak was obtained by measuring the number of scattered neutrons at each rocking angle ($\omega$) with a $80 \times 80$ mm$^2$ area detector. 
	
	Inelastic neutron scattering measurements were performed on the triple-axis neutron spectrometer IN8~\cite{Hiess2006} with the FlatCone detector~\cite{Kempa2006} at the ILL.	A \ymb\ single crystal (mass 1\,g) was initially oriented with the $a$ and $c$ crystal axes  horizontal  to map the spin-wave spectrum in the ($h$\,0\,$l$) scattering plane (see Fig.~\ref{fig:Unitcell}). The crystal was subsequently rotated by $90^{\circ}$ (such that the crystalline $a$ and $b$ axes were in the scattering plane) to access the ($h$\,$k$\,0) plane. Constant-energy maps were measured at various energies, $\Delta E = E_\textrm{i}- E_\textrm{f}$. The outgoing neutron wavevector was fixed at $k_\textrm{f}$ = 3\,\AA$^{-1}$ ($E_\textrm{f} =18.6$ meV) by elastically-bent Si $(111)$ analyzer crystals, and the required energy transfers were set by selecting the incident wavevector, $k_\textrm{i}$, with an incident beam monochromator. For energy transfers $\Delta E \geq 40$\,meV, a PG $(002)$ double-focusing monochromator was used, and for $\Delta E < 40$\,meV an elastically-bent, perfect Si $(111)$ double-focusing monochromator was used.
	
	The array of 31 detectors on the FlatCone device allows for the simultaneous acquisition of scattered intensity along arcs in reciprocal space. By rotating the single crystal about the scattering plane normal, these arcs can sweep out areas in \textbf{k}-space to give reciprocal space maps. 
	
	\section{Results and analysis}
	\begin{figure}[t!]
		\includegraphics[width=0.5\textwidth]{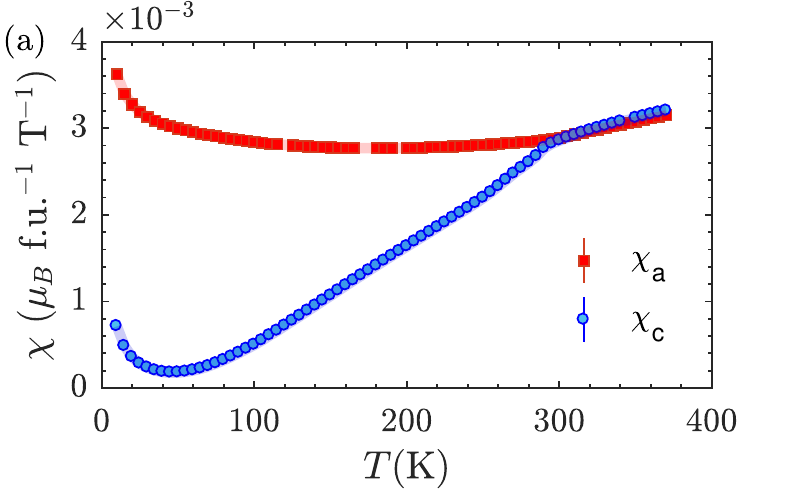}
		\includegraphics[width=0.5\textwidth]{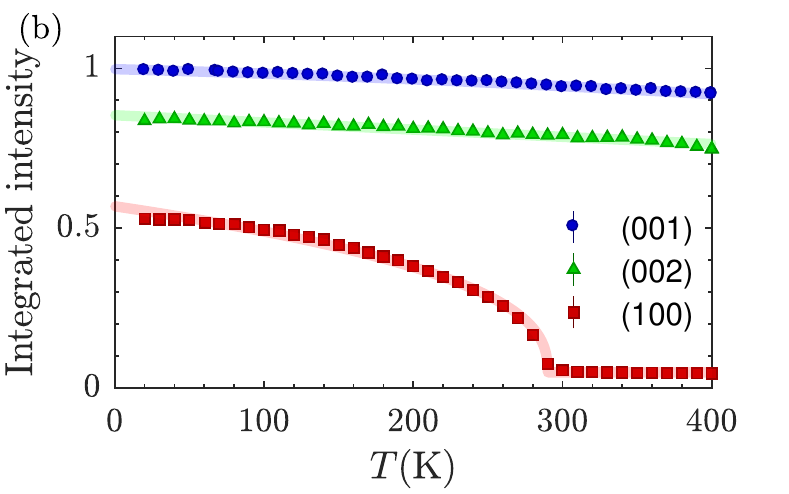}
		\includegraphics[width=0.5\textwidth]{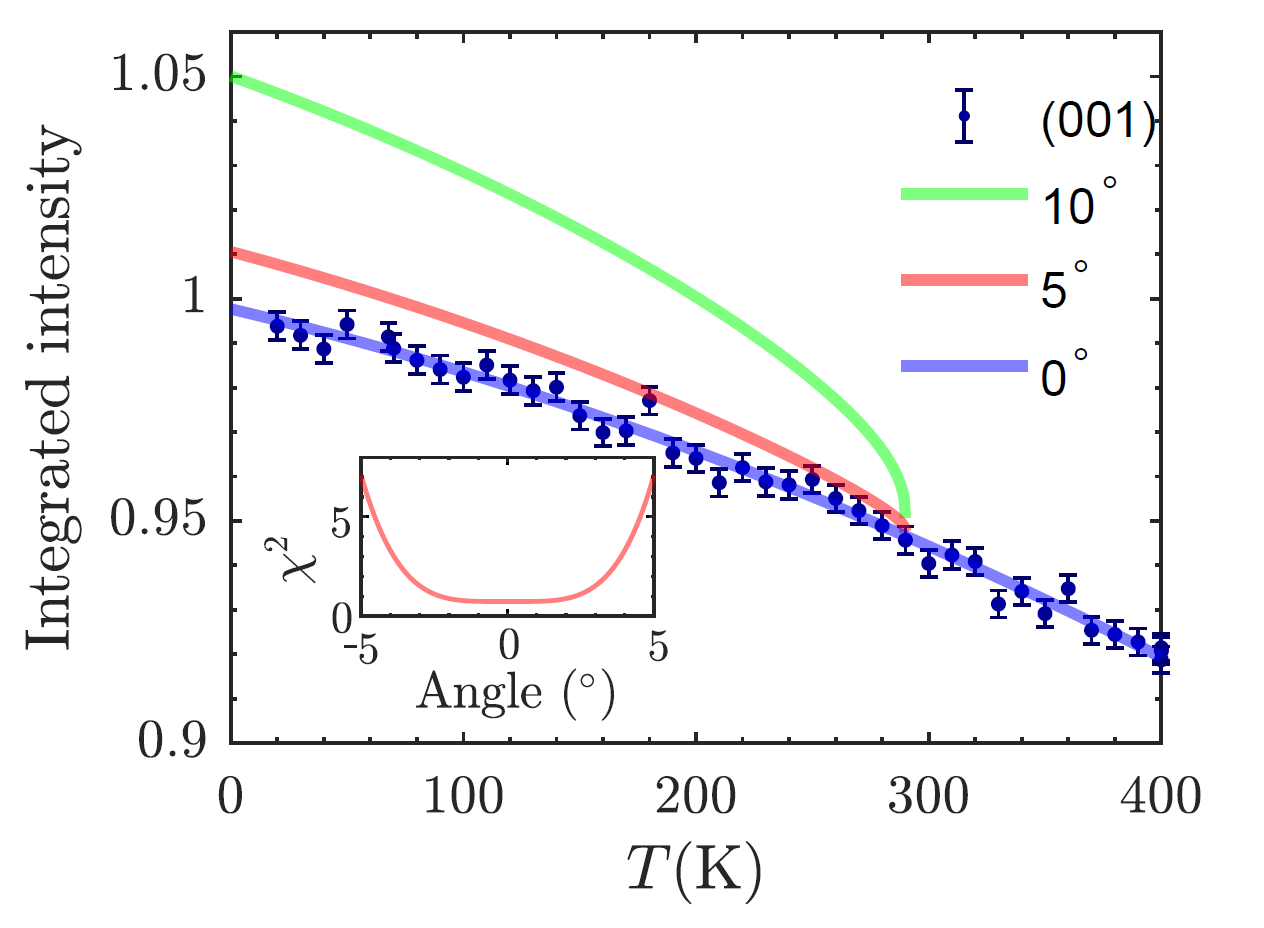}
		\caption{\label{fig:DH3_DL5_YbMnBi2} (a) Temperature dependence of the magnetic susceptibility of YbMnBi$_2$ measured with the field applied along the $a$ and $c$ axes ($\chi_a$ and $\chi_c$, respectively). The single crystal was cooled in zero field and measured in an applied field strength of 1\,T. (b) Temperature dependence of the integrated intensity of the $(001)$, $(002)$ and $(100)$ peaks. The red line is a power law fit to the temperature dependence of the $(100)$ reflection which gives a transition temperature of $T_\textrm{N}$=290(1)\,K. (c) Measured intensity of the $(001)$ peak, together with lines calculated for tilt angles of $0^\circ$, $5^\circ$ and $10^\circ$. The inset shows the variation of the $\chi^2$  with tilt angle. }
	\end{figure}
	The x-ray diffraction patterns of single crystalline \ymb\, obtained from the 6-circle and Laue diffractometers are fully consistent with the $P4/nmm$ space group, with cell parameters $a = 4.4860(13)$\,\AA\,and $c = 10.864(4)$\,\AA\,(Ref.~\onlinecite{YbMnBi2Supp}). Moreover, the small mosaic spread in the diffraction peaks ($< 1.24 ^{\circ}$) points to a high crystalline quality of the flux-grown crystals.
	
	The temperature dependence of the magnetic susceptibility of YbMnBi$_2$, with the field applied parallel to the $a$ and $c$ crystal axes, is shown in Fig.~\ref{fig:DH3_DL5_YbMnBi2}(a). The anomaly in the $\chi_c$ data at $T_{\rm N} \simeq 290$\,K is associated with the onset of AFM order in the Mn$^{2+}$ sublattice. This value for the N\'{e}el temperature is consistent with those reported in earlier studies of YbMnBi$_2$~\cite{Borisenko2015,Wang2016,Liu2017}, as well as the  neutron diffraction data presented in this work (see later). Below $T_\mathrm{N}$, the magnetic susceptibility becomes strongly anisotropic with respect to applied field, where $\chi_a > \chi_c$. This bifurcation of $\chi(T)$ at $T_\textrm{N}$ suggests that the manganese moments, in the ordered phase, are more susceptible to an in-plane field than a field applied along the $c$-axis, in agreement with earlier reports~\cite{Borisenko2015,Wang2016}. At low temperatures (below 50\,K), the susceptibility grows in both field directions. This upturn is likely due to a small concentration of a Mn-containing paramagnetic impurity phase, and is observed in other members of the $A$MnBi$_2$ family ($A$ = Sr, Ca, Ba)~\cite{Guo2014,Li2016,Wang2016b}. 
	
	\subsection{Elastic Neutron Scattering}
	Neutron diffraction data in the temperature range 20 to 400\,K are presented in Fig.~\ref{fig:DH3_DL5_YbMnBi2}(b). As the sample was cooled below $T = 290$\,K, the $(100)$ peak, which is otherwise forbidden in the $P4/nmm$ space group, was observed. This reflection is consistent with a magnetic propagation vector of $\textbf{k} = \textbf{0}$. The onset of this purely magnetic peak at $T_\mathrm{N}$ reveals the incipient AFM order of the Mn$^{2+}$ sublattice. The temperature dependence of the integrated peak intensity fits very well to a power law, $I_{obs}\propto |T_\mathrm{N}-T|^{2\beta}$ , with critical exponent $\beta$ = 0.38(2), consistent with the 3D Heisenberg universality class. 
	
	The predicted canting of the Mn$^{2+}$ moments away from the $c$--axis~\cite{Borisenko2015,Chaudhuri2017,Chinotti2016} should produce a small $ab$--plane ferromagnetic component. Given that magnetic neutron scattering is sensitive to the component of the ordered moment perpendicular to the scattering vector \textbf{Q},\cite{Squires2012} we can isolate this small in-plane component by studying the intensity of reflections with $\textbf{Q} \parallel c$. If there were an in-plane ferromagnetic component then the intensity of $(00l)$ peaks should increase on cooling below $T_{\rm N}$, as was observed in a sister compound SrMnSb$_2$~\cite{Liu2017b}, where a small in-plane ferromagnetic contribution to the nuclear peak was reported\footnote{Note that the $a$ and $c$ axis in Ref.~\onlinecite{Liu2017b} are interchanged with respect to those defined in the present work. SrMnBi$_2$ suffers from an off stoichiometry and is better described by Sr$_{1-y}$Mn$_{1-z}$Sb$_2$ $(y, z < 0.1)$.}.
	
	To minimize the reduction of the scattered intensity due to the magnetic form factor of Mn$^{2+}$, we studied the reflections with the smallest \textbf{Q}, namely the $(001)$ and $(002)$ peaks, as shown in Fig.~\ref{fig:DH3_DL5_YbMnBi2}(b). We observe no discernible change in the integrated intensity of these peaks apart from the gradual increase with decreasing temperature which can be attributed to the Debye--Waller factor. 
	
In Fig.~\ref{fig:DH3_DL5_YbMnBi2}(c) we show the intensity of the $(001)$ peak on a magnified scale, together with lines calculated  assuming tilt angles of $0^\circ$, 5$^\circ$ and 10$^\circ$. The $0^\circ$ curve is a quadratic fit to the data, and the other two curves are obtained by adding the calculated magnetic intensity of the $(001)$ peak to the $0^\circ$ curve based on the measured intensity of the $(100)$ peak. We also calculated the variation of the $\chi^2$ goodness-of-fit statistic as a continuous function of tilt angle, see inset to Fig.~\ref{fig:DH3_DL5_YbMnBi2}(c). From the $\chi^2$ distribution, we find that the probability of a tilt angle greater than $3^\circ$ is only 5\%. 
 
 These results imply that the ordered moments in YbMnBi$_2$ are collinear and aligned along the $c$-axis to within $3^\circ$ at a $95\%$ confidence level. Hence, a 10$^\circ$ canting of Mn$^{2+}$ moments away from the $c$-axis, as required to create the Weyl nodes, can be excluded. 
	%This finding is in agreement with a recent neutron diffraction experiment on single crystalline YbMnBi$_2$, where reflections in the ($h$\,0\,$l$) scattering plane were studied at $T=4$\,K and $T=300$\,K~\cite{Zaliznyak2017}. As polarized neutrons were used, Zaliznyak \etal were able to distinguish between the magnetic and nuclear contributions to the observed peaks. Hence, if there were a small magnetic contribution to the nuclear reflections, which might arise from the canting of the Mn$^{2+}$ moments, this technique would be sensitive to it. They reported no perceptible  
	%, observable, perceivable, distinguishable, recognizable, identifiable 
	%change in the magnetic component in the $(00l)$ family of reflections at the two temperatures, and concluded that the moments point along the $c$-axis, in agreement with the findings of our work.}
	
	\subsection{Inelastic Neutron Scattering}

	\begin{figure}[t!]
		\includegraphics[width=0.45\textwidth]{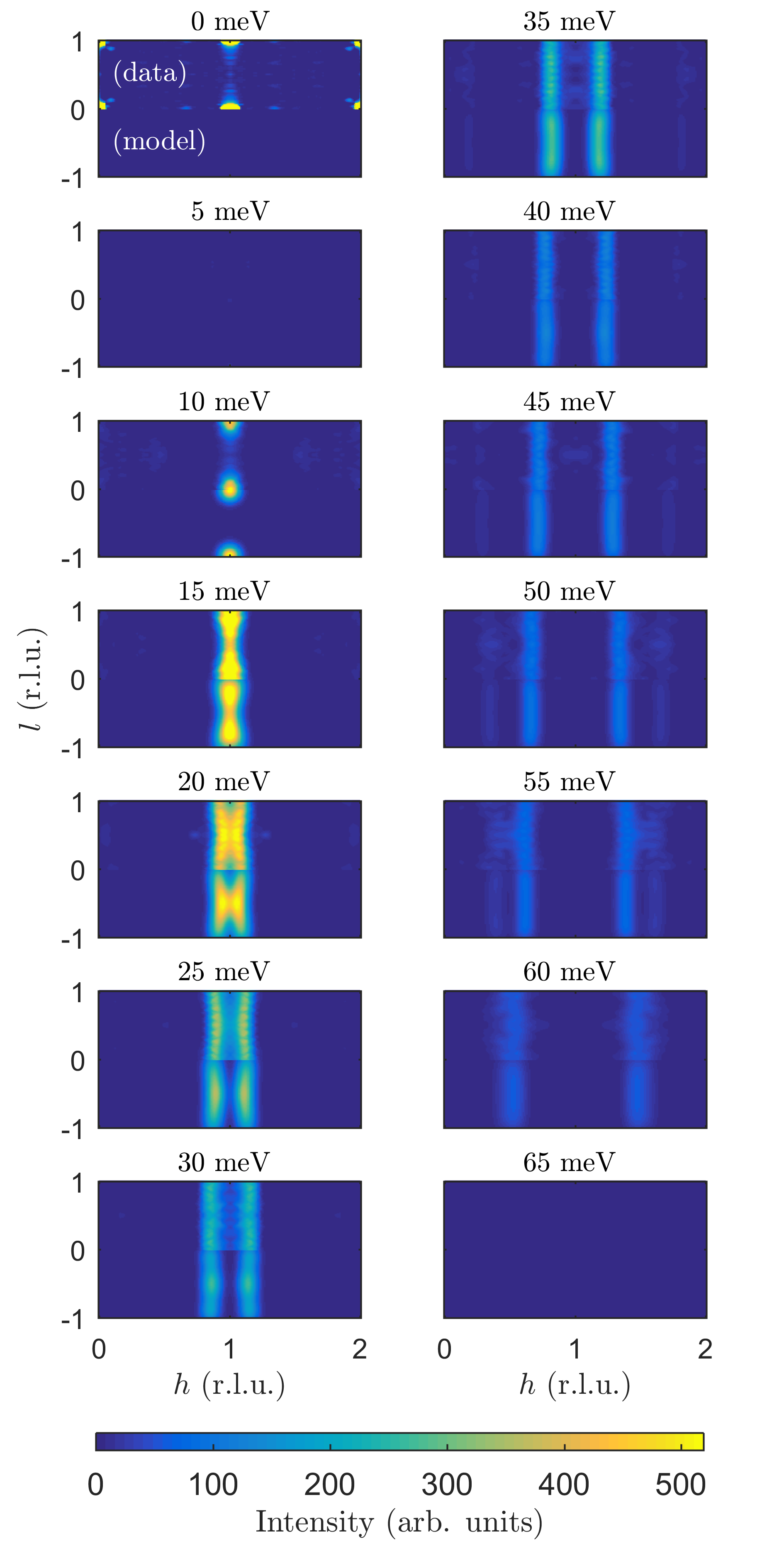}
		\caption{\label{fig:DA7_YbMnBi2_SpinW_H0L} Constant-energy maps in the $(h0l)$ plane in reciprocal space, illustrated in Fig.~\ref{fig:Unitcell}(b), at various $\Delta E$, plotted in reduced lattice units (r.l.u.). In each panel, the top and bottom half correspond to the data and model, respectively.}
	\end{figure} 
	
	\begin{figure}[t!]
	\includegraphics[width=0.45\textwidth]{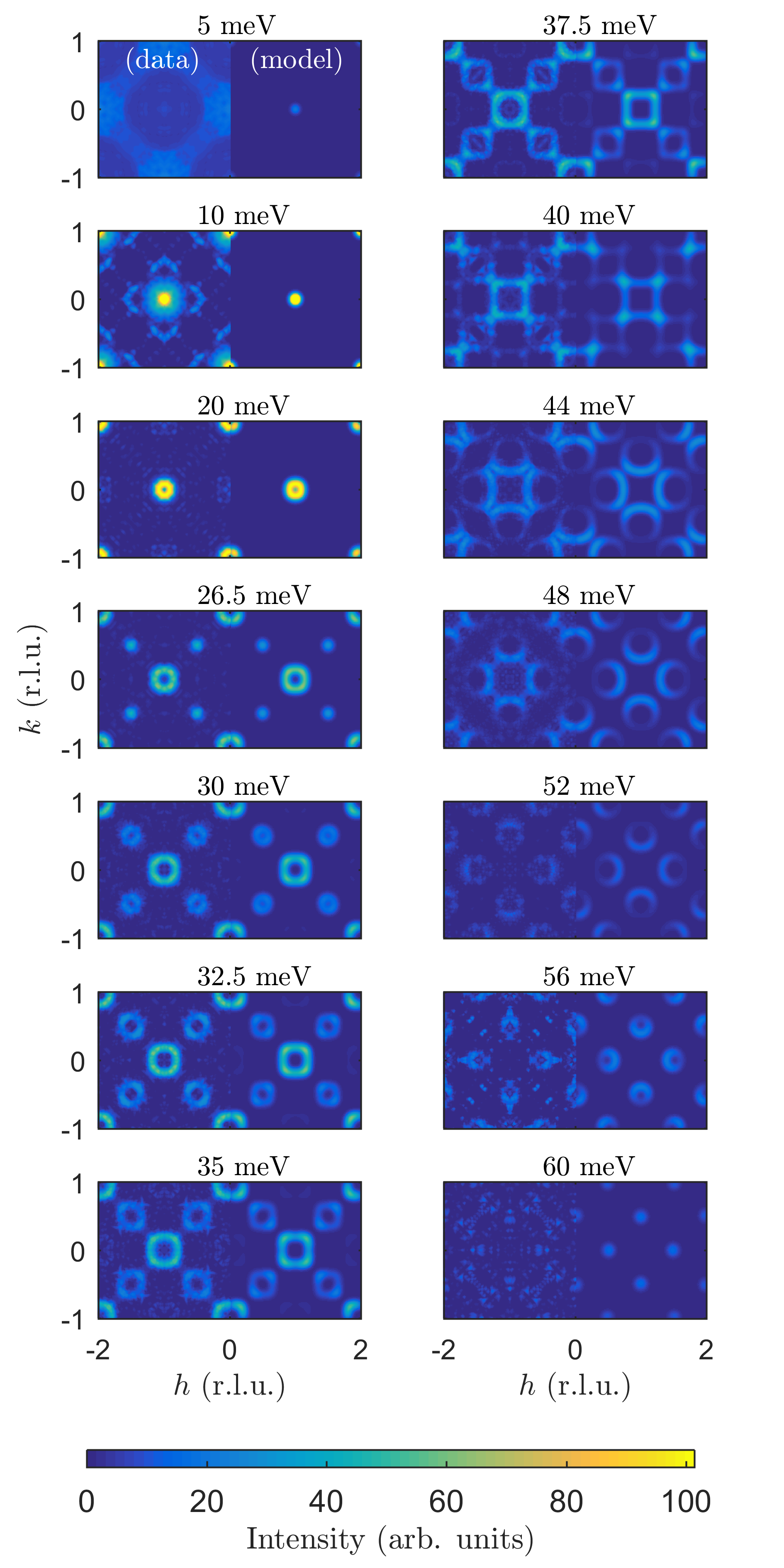}
	\caption{\label{fig:DA7_YbMnBi2_SpinW_HK0} Constant-energy maps in the $(hk0)$ plane in reciprocal space, illustrated in Fig.~\ref{fig:Unitcell}(b), at various $\Delta E$. In each panel, the left and right half correspond to the data and model, respectively.}
	\end{figure}
	Constant-energy maps of the scattering intensity recorded in the $(h0l)$ and $(hk0)$ reciprocal lattice planes at various energy transfers, $\Delta E$, are shown in Figs.~\ref{fig:DA7_YbMnBi2_SpinW_H0L} and~\ref{fig:DA7_YbMnBi2_SpinW_HK0}, respectively. We discuss the data from the different scattering planes in turn, starting with the $(h0l)$ data, which appears in the top half of each panel in Fig.~\ref{fig:DA7_YbMnBi2_SpinW_H0L}.

	We find the lowest energy spin-wave mode at the $\Gamma$ point, with an energy gap of $\Delta E \simeq 10$\,meV. This gap is caused by the magnetic anisotropy which favors spin alignment along the $c$ axis. At $\Delta E$= 20 meV, we find pinch points in the magnon spectrum at the high symmetry point $Z$, that is, halfway between $\Gamma$ points in adjacent Brillouin zones along $l$.  These pinch points form as a result of the dispersion along the $c$-axis. For $\Delta E \geq 30$ meV, the magnon dispersion along $l$ goes away, and the intensity becomes independent of $l$. In other words, the Mn spin dynamics becomes two-dimensional.
	%at these energies, the phases of the spin-wave modes propagating in adjacent $ab$ planes become uncorrelated.  This indicates that the inter-plane exchange coupling of the ordered Mn$^{2+}$ moments is relatively weak compared to the coupling within the $ab$--plane. Moreover, given the large separation of the manganese ions along the $c$ axis compared to that within the plane, this also suggests that the spin-wave spectrum of YbMnBi$_2$, at these energy transfers, can be described by ordered Mn$^{2+}$ moments on an isolated 2D square lattice. %Moreover, given that $c/a \sim 2.5$, there are 4 nearest-neighbour in
	The spectrum reaches a maximum along the $R-X-R$ high symmetry line at $\Delta E$= 60 meV.

	We now turn to the reciprocal space maps in the $(hk0)$ scattering plane at various energy transfers, which correspond to the left half of each panel in Fig.~\ref{fig:DA7_YbMnBi2_SpinW_HK0}.  Just as in the $(h0l)$  plane, we observe the lowest energy excitations at the $\Gamma$ point in the Brillouin zone at $\Delta E = 10$\,meV. For $10\leq \Delta E \leq 26.5$\,meV, the spectrum develops into rings centered at $\Gamma$, which is characteristic of isotropically dispersing spin waves in the $ab$ plane. 
	%At these small wavevectors ($\abs\textbf{k}} = 2\pi/\lambda$) the in-plane exchange coupling between Mn$^{2+}$ moments appear isotropic. This arises because the magnon excitations at large wavelengths ($ \lambda \gg a $) are insensitive to the periodicity of the square lattice. When the wavelength of the spin-wave becomes the same order of magnitude as the lattice spacing, that is $\lambda \sim a$, the magnon spectrum becomes progressively less isotropic, as shown in the $\Delta E \geq 30$\,meV  data. Here the spin-wave excitations become sensitive to the different exchange coupling strength between the $ab$-plane nearest and next-nearest neighbor interactions. This breaks the rotational symmetry about $\Gamma$ and spin-wave spectrum adopts the lower $\bar{4}m2$ symmetry of the Mn$^{2+}$ ions in the $ab$--plane. 
	At $\Delta E$= 26.5 meV, we observe a saddle in the spin-wave spectrum appearing at the high symmetry point $M$. 
	%Eventually, at $\Delta E \simeq 40$\,meV, the magnon modes emanating from $M$ and $\Gamma$ meet at pinch points in the constant-energy maps at $\sim 1/4$ and $\sim3/4$ along the $\Gamma-M-\Gamma$ high symmetry line. Here the spectrum develops into a square network which is relatively linear in $h$ and $k$. {\color{red} I don't think it is necessary to describe the evolution in such detail.}
	The maximum in the dispersion is once again found at the $X$ point, at $\Delta E \leq 60$\,meV.
	%At larger energy transfers, $44 \leq \Delta E \leq 60$ meV, the in-plane spin wave modes propagating along $a$ eventually terminates at the high symmetry point $X$, in agreement with the $(h0l)$  data.

\begin{figure}[t]
		\includegraphics[width=0.5\textwidth]{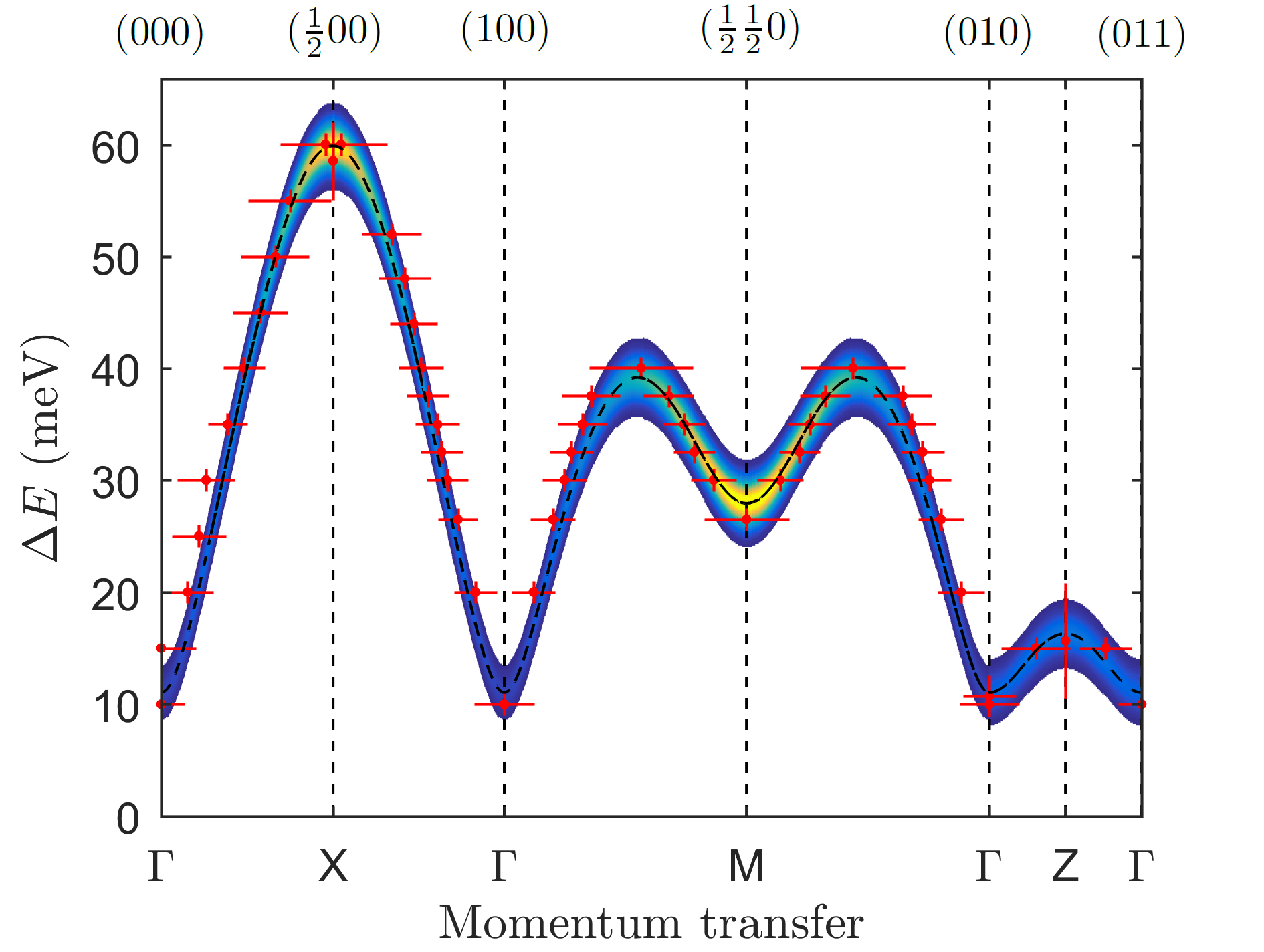}
		\caption{\label{fig:DA6_YbMnBid2_SpinW} The observed and calculated spin-wave spectrum of the Mn spins in \ymb\, along  high symmetry directions, as defined in Fig.~\ref{fig:Unitcell}(b). The calculated magnon spectrum is in good agreement with the measured spin-wave dispersion (red markers), which was obtained from constant-energy cuts through the intensity maps in the $(h0l)$ and $(hk0)$ planes.}
	\end{figure}

	To obtain the spin-wave dispersion, cuts were made along the $Z-\Gamma-X$ and $M-\Gamma-X$ high symmetry lines [see Fig.~\ref{fig:Unitcell}(b)] through the measured intensity maps in the $(h0l)$ and $(hk0)$ planes, respectively. The intensity in cuts at various $\Delta E$ was fitted with peak functions to identify the magnon wavevectors for each $\Delta E$. In Fig.~\ref{fig:DA6_YbMnBid2_SpinW} we present the measured spin-wave dispersion determined this way. 
	
	In order to model the observed magnon spectrum we employed the effective spin Hamiltonian 
		\begin{align}
	{\mathcal H} = \sum_{i,j}J_{ij} \textbf{S}_i \cdot \textbf{S}_j - \sum_i D(S^z_i)^2,
	\label{eq:SpinWeqn}
	\end{align}
	where $J_{ij}$ is the (isotropic) exchange between Mn spins $\textbf{S}_i$ and $\textbf{S}_j$ on sites $i$ and $j$, and $D$ is a single-ion anisotropy parameter making the $c$ axis an easy axis. In the first summation, we include first and second nearest neighbors in the $ab$ plane ($J_1$ and $J_2$), and nearest neighbors along the $c$ axis ($J_c$). We used linear spin-wave theory as implemented in the SpinW software\cite{Toth2015} to calculate the magnon spectrum.

	By fitting the linear spin-wave model to the measured dispersion we find values for the parameters $SJ_1$ = 22.6(5)\,meV, $SJ_2$ = 7.8(5)\,meV, $SJ_c = -0.13(5)$\,meV and $SD$ = 0.37(4) meV  (see Supplemental Material for details\cite{YbMnBi2Supp}), where $S$ is the spin quantum number, which for Mn$^{2+}$ is $S=5/2$. Based on these parameters, we present the calculated constant-energy intensity maps in the $(h0l)$ and $(hk0)$ planes on the lower and right halves of the panels in Figs.~\ref{fig:DA7_YbMnBi2_SpinW_H0L} and ~\ref{fig:DA7_YbMnBi2_SpinW_HK0}, respectively, and we plot the calculated magnon spectrum along high symmetry directions in Fig.~\ref{fig:DA6_YbMnBid2_SpinW}. Overall, we find that the calculated spectrum agrees very well with the data.
	
	\section{Discussion}
	
	As neutron diffraction probes the entire volume of the sample, our results rule out the possibility of magnetically-induced Weyl nodes in the bulk of YbMnBi$_2$. On the other hand, neutron diffraction would not be sensitive to a canting of the magnetic moments at the surface of the sample. Such a canting, if present, would reconcile the results of the present study with the work by Borisenko \textit{et al.}~\cite{Borisenko2015}. 
	
	In \ymb, the spontaneous magnetic order in the Mn sublattice coexists with massless quasiparticle excitations arising from the Bi square net. Armed with the best-fit parameters of the linear spin-wave model, we are now in the position to address whether the magnon spectrum in \ymb\, differs in any detectable way compared with other related systems. %by a coupling between the two phenomena which may reflect the presence of Weyl nodes near $E_\mathrm{F}$. 
 For instance, one might expect to see differences in the inter-layer exchange coupling  parameter $J_c$ if the conducting states on the Bi layers were very unusual in \ymb\,.
	%This, for instance, may be manifest in the strength of the interlayer coupling between Mn$^{2+}$ ions which is quantified by the magnetic exchange parameter $SJ_c$. 
	
 To elucidate this, we compare the fitted spin-wave model parameters obtained in this work with those of CaMnBi$_2$, which is isostructural to \ymb\,. CaMnBi$_2$ possesses a near identical N\'{e}el temperature to \ymb\, of $T_\mathrm{N}$ = 290\,K,~\cite{Rahn2017,Guo2014} and is predicted to be a Dirac semimetal.~\cite{Feng2014,WangKF2012,Zhang2012} 
	%This means that the electronic bands in CaMnBi$_2$ are two-fold degenerate. 
	%as TRS is preserved compared to \ymb\, where this degeneracy is predicted to be lifted due to the purported spontaneous TRS breaking~\cite{Borisenko2015}. 
	Using the same Hamiltonian (\ref{eq:SpinWeqn}), the three magnetic exchange parameters in CaMnBi$_2$ were found to be $SJ_1$ = 23.4(6)\,meV, $SJ_2$ = 7.9(5)\,meV and $SJ_c$ = $-0.10(5)$\,meV,~\cite{Rahn2017} which are the same as those of \ymb\, to within experimental error.The anisotropy parameter for  CaMnBi$_2$,  $SD = 0.18(3)$\,meV, is about half that for YbMnBi$_2$, which reflects that the energy gap at $\Gamma$ is slightly smaller in CaMnBi$_2$ than in YbMnBi$_2$. These results demonstrate that the magnon spectrum of \ymb\, does not show any anomalous behavior relative to that of CaMnBi$_2$.
	
	More broadly, this suggests that replacing the divalent alkali-earth metal Ca$^{2+}$ on the $A$ site of $A$MnBi$_2$ with the rare-earth Yb$^{2+}$ ion does not significantly enhance the coupling between the magnetism in the octahedral MnBi$_4$ layers and the charge carriers in the Bi square net. This is despite the fact that the $A$ atom is situated along the direct exchange path between the Mn and Bi atoms. In a recent review of the wider $A$Mn$Pn_2$ family of compounds,  Klemenz \textit{et al.}~\cite{Klemenz2019} suggested another route to enhance the coupling between magnetism and the topological charge carriers, namely to have a magnetic ion on the $A$ site (like Eu$^{2+}$) rather than non--magnetic ions such as Ca$^{2+}$, Sr$^{2+}$, Ba$^{2+}$ and Yb$^{2+}$. This was prompted by the fact that the $A$ site atom is in closer proximity to the square Bi compared to the Mn$^{2+}$ ion and might lead to a greater orbital overlap and thus magnetic exchange interaction. In fact, this was considered in Refs.~\onlinecite{Chinotti2016,Borisenko2015}, where the electronic structure and optical properties of EuMnBi$_2$ and \ymb\, were compared. The divalent rare-earth ions on the $A$ site of both $A$MnBi$_2$ compounds have comparable ionic radius and very similar relative positions to the Bi square layer, but with the difference that Eu$^{2+}$ has half-filled $4f$ orbitals compared to the fully-filled case for Yb$^{2+}$. This leads to a large pure-spin magnetic moment of 7$\mu_\mathrm{B}$ on the $A$ site of EuMnBi$_2$, and a non-magnetic ion  on the $A$ site of \ymb\,. These studies demonstrate a marked increase in %$f-p$ 	exchange
	coupling between magnetism and the topological charge carriers in EuMnBi$_2$ compared to that in \ymb, which is consistent with magnetotransport studies~\cite{Borisenko2015,May2014,Masuda2016,Masuda2018,Wang2016}. This suggests that in EuMnBi$_2$, compared to \ymb, a greater coupling of magnetism to the pnictide square net can be achieved with magnetic species on the $A$ site, which for the extended $A$Mn$Pn_2$ (or 112-pnictide) family, is closer to the pnictide layer compared to Mn.

	%and $SD$ = 0.18(3) meV) is
	%As discussed earlier, this might arise from the coupling between the predicted Weyl fermions in the square Bi net and the Mn sublattice that may lead to changes in the spin dynamics.

	Finally, it is instructive to compare the physical properties of \ymb\, with that of YbMnSb$_2$, which is isostructural to YbMnBi$_2$~\cite{Wang2018,Kealhofer2018} and also exhibits Mn AFM order with a similar magnetic ordering temperature of $T_\mathrm{N}$\,=\,345\,K. A comparison of the band structures of the two 112 pnictides reveal a greater extent of inversion in the conduction and valence bands in YbMnBi$_2$, with several band crossings at $E_{\rm F}$ as shown Refs.~\onlinecite{Borisenko2015,Chaudhuri2017}, compared to that in YbMnSb$_2$.~\cite{Kealhofer2018} Moreover, the Shubnikov–de Haas (SdH) oscillation of the magneto--transport in both compounds reveals that the effective mass of the charge carriers in \ymb\, ($m^\ast_c \sim 0.24\,m_e$~\cite{Liu2017}) is approximately twice that of YbMnSb$_2$ as reported in Refs.~\onlinecite{Kealhofer2018,Wang2018}.
	
	These features can be understood from the relative sizes of the spin--orbit coupling (SOC) in the pnictide square conducting layers, which is significantly larger in \ymb\, as Bi is $\sim 1.7$ times heavier than Sb. Given that the linear band crossing along the $\Gamma$--$M$ high symmetry line is not protected by symmetry, the doubly--degenerate pnictide (Sb 5$p$ or Bi 6$p$) bands hybridize and give rise to an avoided Dirac crossing. As such, the stronger SOC in \ymb\, produces a larger energy gap in the electronic bands, resulting in a heavier effective mass of the charge carriers compared to that in YbMnSb$_2$. This is consistent with the work in Ref.~\onlinecite{Liu2016}, which explored the effect of the masses of pnictides on the physical properties of BaMn$Pn_2$ ($Pn$ = Sb, Bi). In that work, Liu \textit{et al.} also proposed that a more suitable platform to realize massless Dirac fermions is in replacing Bi with lighter elements in the same group. This demonstrates that the 112 pnictide family of compounds offers strong tunability of the effective mass of the charge carriers from the size of the SOC.
	
	%\textbf{One disadvantage of AMnBi$_2$ as Dirac semimetals is that the strong spin orbit coupling (SOC) due to heavy Bi atoms opens gap } at Dirac nodes, leading to massive Dirac electrons. For instance, in \ymb the SOC-induced gap at the Dirac node is about 4 meV and the effective mass of Dirac fermions estimated from the analyses of Shubnikov-de Haas (SdH) oscillations is 0.29m0 (m0, the mass of free electron) 23, one possible route to realize massless Dirac fermions in AMnBi2-type material is to replace Bi with other lighter main group elements such as Sb and Sn, whose SOC effect is much weaker.
	\section{Conclusion}
	We have presented the magnetic structure and magnon spectrum of the candidate Weyl semimetal \ymb.  The $(0\,0\,l)$ family of nuclear reflections does not display any additional magnetic contribution below $T_{\rm N}$, and this rules out the mechanism for creation of Weyl nodes via TRS-breaking through canting of the Mn spins. Hence, we demonstrate that bulk \ymb\, is a Dirac semimetal rather than a host for the WSM state. We have not ruled out the possibility of spin canting at the surface, which could reconcile the present results with those of Ref.~\onlinecite{Borisenko2015}. The lack of any anomalous features in the magnon spectrum implies a weak coupling between magnetism and the topological charge carriers. \ymb\, belongs to the wider $A$Mn$Pn_2$ family of compounds which are currently attracting strong interest owing to its strong potential for spintronic applications. We hope that the understanding of \ymb\, achieved here will contribute to the development of strategies for enhancing the exchange coupling between charge transport and magnetism, and for reducing the effective mass of the quasiparticles.
	\begin{acknowledgments}
		The authors wish to thank D. Prabhakaran and F. Charpenay for technical assistance, and M. Newport for fabricating the Al mount used in the INS experiment. We are also grateful to M. C. Rahn and P. Steffens for the data analysis software, P. Manuel and D. D. Khalyavin for help with preliminary neutron studies on WISH, ISIS (beamtime RB1720113), M. Gutmann for checking the single crystal quality of \ymb\, on SXD, ISIS and N. Qureshi for orienting the crystal for the INS experiment on OrientExpress~\cite{Ouladdiaf2006}, ILL (beamtime EASY-365). The D10 and IN8 experiment numbers were DIR-159 and 4-01-1572 ~\cite{Boothroyd2018} respectively. This work was supported by the U.K. Engineering and Physical Sciences Research Council, Grant Nos. EP/N034872/1 and EP/M020517/1, the Natural Science Foundation of Shanghai (Grant No. 17ZR1443300), the Shanghai Pujiang Program (Grant No. 17PJ1406200), the National Key Research and Development Program of China (Grant No. 2017YFA0302901), the Beijing Natural Science Foundation (Grant No. Z180008) and the K. C. Wong Education Foundation (Grant No. GJTD-2018-01). J.-R. Soh acknowledges support from the Singapore National Science Scholarship, Agency for Science Technology and Research.

	\end{acknowledgments}
	
	\nocite{*}
	\bibliography{main}
\end{document}